\documentclass[useAMS,usenatbib]{mn2e}

\usepackage[T1]{fontenc} 
\usepackage{aecompl}
\usepackage{amssymb}
\usepackage[fleqn]{amsmath}
\usepackage{color}
\usepackage[bookmarks,bookmarksopen,colorlinks,linkcolor={blue},citecolor={red},urlcolor={green}]{hyperref}
\usepackage{graphicx}
\usepackage{times}
\usepackage{epsfig}
\usepackage{hyperref}

\def\be{\begin{equation}}
\def\ee{\end{equation}}

\def\gtsima{$\; \buildrel > \over \sim \;$}
\def\ltsima{$\; \buildrel < \over \sim \;$}
\def\prosima{$\; \buildrel \propto \over \sim \;$}
\def\gsim{\lower.5ex\hbox{\gtsima}}
\def\lsim{\lower.5ex\hbox{\ltsima}}
\def\simgt{\lower.5ex\hbox{\gtsima}}
\def\simlt{\lower.5ex\hbox{\ltsima}}
\def\simpr{\lower.5ex\hbox{\prosima}}

\newcommand{\beq}{\begin{equation}}
\newcommand{\eq}{\end{equation}}
\newcommand{\bear}{\begin{eqnarray}}
\newcommand{\ear}{\end{eqnarray}}

\newcommand{\sdz}{$\sigma^2(z)$\ }

\def\beqn{\begin{eqnarray}}
\def\eeqn{\end{eqnarray}}
\def\pnl{{P}_{{\!\textrm{\tiny NL}}}}
\def\dnl{\Delta_{\text{\scriptsize NL}}}

\def\Scr{\Sigma_{\mathrm{crit}}}

\title[Constraining WDM with high-$z$ supernova lensing]{Constraining warm dark matter with high-$z$ supernova lensing}

\author[S. Pandolfi, C. Evoli, A. Ferrara, F. Villaescusa-Navarro]
{S.~Pandolfi$^{1}$, C.~Evoli$^{2}$, A.~Ferrara$^{3}$, F.~Villaescusa-Navarro$^{4}$ \\
$^{1}$ Dark  Cosmology  Centre,  Niels  Bohr  Institute,  University  of  Copenhagen,  Juliane  Maries  Vej  30,  DK-2100  Copenhagen,Denmark\\
$^{2}$ II. Institut f\"ur Theoretische Physik, Universit\"at Hamburg, Luruper Chaussee 149, D-22761 Hamburg, Germany\\
$^{3}$Scuola Normale Superiore, Piazza dei Cavalieri 7, I-56126 Pisa, Italy\\
$^{4}$INAF - Osservatorio Astronomico di Trieste, Via Tiepolo 11, I-34143 Trieste, Italy\\
}

\begin{document}
\maketitle
\label{firstpage}

\begin{abstract}
We propose a new method to constrain the warm dark matter (WDM) particle mass, $m_\chi$, based on the counts of multiply imaged, distant supernovae (SN) produced by strong lensing by intervening cosmological matter fluctuations. The counts are very sensitive to the WDM particle mass, assumed here to be $m_\chi=1, 1.5, 2$ keV. We use the analytic approach developed by Das \& Ostriker to compute the probability density function of the cold dark matter (CDM) convergence ($\kappa$) on the lens plane; such method has been extensively tested against numerical simulations. 
We have extended this method generalizing it to the WDM case, after testing it against WDM $N$-body simulations. Using the observed cosmic star formation history we compute the probability for a distant SN to undergo a strong lensing event in different cosmologies. A minimum observing time of 2 yr (5 yr) is required for a future 100 square degrees survey reaching $z \approx 4$ ($z \approx 3$) to disentangle at 2$\sigma$ a WDM ($m_\chi=1$ keV) model from the standard CDM scenario. 
Our method is not affected by any astrophysical uncertainty (such as baryonic physics effects), and, in principle, it does not require any particular dedicated survey strategy, as it may come as a byproduct of a future SN survey.
\end{abstract}


\section{Introduction}
\label{sec:intro}

The standard cosmological model postulates that about 25\% of the energy content of the Universe is in the form of non-baryonic cold dark matter (CDM). According to this scenario,  the growth of structures proceeds bottom-up in a hierarchical manner: small structures form first and they later merge into larger structures. Such scheme is a consequence of the ``coldness'' of dark matter, and has been successful in explaining many different cosmological observations~\citep{Ade:2013}. 
Quite uncomfortably, though, there are known problems in such framework that mostly plague small scales: the standard $\Lambda$CDM model seems to predict too much power on small scales, and therefore too many low-mass structures as, e.g., galaxy satellites. This is the so-called ``missing satellite problem'': $N$-body simulations predict about 10 times more satellite galaxies around the Milky Way than observed \citep[see, e.g.,][]{Moore:1999}. 
A second problem is that $\Lambda$CDM simulations show Galactic dark matter haloes with density profiles too centrally concentrated compared to those derived from dynamical data of the Milky Way satellites~\citep[see, e.g.,][]{BoylanKolchin:2011}. 
Moreover, simulations predict a power-law density profile $\rho \sim r^{-1}$ for dark matter-dominated dwarf-galaxies inner profiles, where observations show that they have shallower density cores~\citep[see, e.g.,][]{Maccio:2012,Schneider:2012}. 
Finally, the properties and the distribution of observed voids do not seem to be in agreement with what one would expect from a $\Lambda$CDM model, as the theory again predicts too many structures on small scales~\citep[see, e.g.,][]{Tikhonov:2009}.

These discrepancies could be resolved by complex astrophysical solutions that could modify the clustering properties of baryons: radiative feedback/photoevaporation by an ultraviolet background, or mechanical feedback from supernovae (SNe) or active galactic nuclei (AGN) winds may help suppressing star formation in small satellite halos~\citep{Governato:2006} 
and make their density profiles flatter~\citep{deSouza:2010}. 
However, there is no clear consensus on a single astrophysical solution to the CDM problems, and an ad hoc fine tuning is required to match all the observations. 
Ideally, one would like to identify a physical process able to suppress  small-scale fluctuations, without affecting larger scales. 

As particle free-streaming due to thermal motions smears out fluctuations whose scale is shorter than the corresponding free-streaming length and prevents gravitational collapse at those scales to occur, it looks as a promising avenue. 
Therefore, in order to reconcile theory with observations, the simplest solution is to replace (completely or in part) CDM with a warm dark matter (WDM) component. As dark matter candidates are classified according to their velocity dispersion, a WDM particle has intermediate streaming properties between ``hot'' and  ``cold'' dark matter candidates. Prototypical hot dark matter particles are neutrinos, which have been quickly rejected as Dark Matter candidates as their large velocity dispersion, and thus free-streaming length, would produce an inverted top-down formation hierarchy, in conflict with observations. On the contrary, a CDM candidates have a free-streaming length so short to be negligible for structure formation.  
Although the theoretical study of the WDM scenario is difficult as it requires $N$-body simulations able to resolve highly non-linear scales~\citep[see, e.g., the discussion in][]{Viel:2005},
 the WDM case has gained renewed interest in the community as a competitive and viable solution for the $\Lambda$CDM unsolved issues we mentioned above
~\citep[see, e.g.,][]{Angulo:2013}. 

If the WDM particles are thermal relics (i.e. at some point they were in equilibrium with the primordial heat bath), they have a simple thermal history and we can easily relate their free-streaming length 
with their mass. This is given by:
\begin{equation}
R_{\rm fs} \approx 0.11\left(\frac{\Omega_{\rm \chi}h^2}{0.15}\right)^{1/3}\left(\frac{m_{\rm \chi}}{\rm keV} \right)^{-4/3} \rm Mpc,
\end{equation}
where $\Omega_{\rm \chi}$ is the total energy density contained in WDM particles relative to the critical density, 
$h$ is the Hubble constant in units of 100 $\rm km \,s^{-1} Mpc^{-1}$ and  $m_{\rm \chi}$ is the WDM particle mass 
(see, for a precise derivation,~\citealt{Bode:2000}; for a pedagogical explanation of the free-streaming length, see \citealt{Lesgourgues:2012}).

Many WDM candidate particles have been proposed. Among them, keV sterile neutrinos are very popular and they have long been discussed and studied~\citep{Dodelson:1994}; light gravitinos~\citep{Nowakowski:1995}, and more recently keV axions~\citep{Conlon:2013} have received some attention.
Different observables have been used to put constraints on the mass of the WDM particle.
Lower limits on the mass of the WDM particles has been established using a variety of observables: Lyman-$\alpha$ forest data suggest that the WDM particle mass should be $m_{\chi}\geq 1$ keV~\citep{Viel:2008}; methods using the stellar mass function and the Tully-Fisher relation~\citep{Kang:2013} 
and cosmic reionization data~\citep{Schultz:2014} yield $m_{\chi}\geq 0.75$ keV; to solve the problems related to the galaxy brightness and stellar mass distribution plaguing CDM discussed above, a WDM particle with a mass of $m_{\chi}\approx 1$~keV seems to represent a viable solution ~\citep{Menci:2013}.
The most recent constrains on WDM particle mass come from the number density of high-redshift ($z\approx 10$) galaxies found by Cluster Lensing And Supernova survey with Hubble (CLASH)
~\citep{Postman:2012}, 
for which $m_{\chi}>1$ keV at 99\% confidence level (c.l.)~\citep{Pacucci:2013}. 
The analysis of the Ly$\alpha$ flux power spectrum of distant ($z>4$) quasars yields the lower limit $m_{\chi}> 3.3$ keV at 2$\sigma$~\citep{Viel:2013}. 
For other methods which have been used to put constraints on the WDM particle mass see, for example,~\citet{Viel:2012} and references therein.
However, these methods are prone to astrophysical uncertainties, such as, for example, the accurate inclusion of baryonic physics in the data analysis and interpretation, which may bias the derived WDM particle mass value.
Moreover, these constraints are model dependent, and may differ if the WDM particle it is not a thermal relic.

\begin{figure}
\centering
\includegraphics[width=8.5cm]{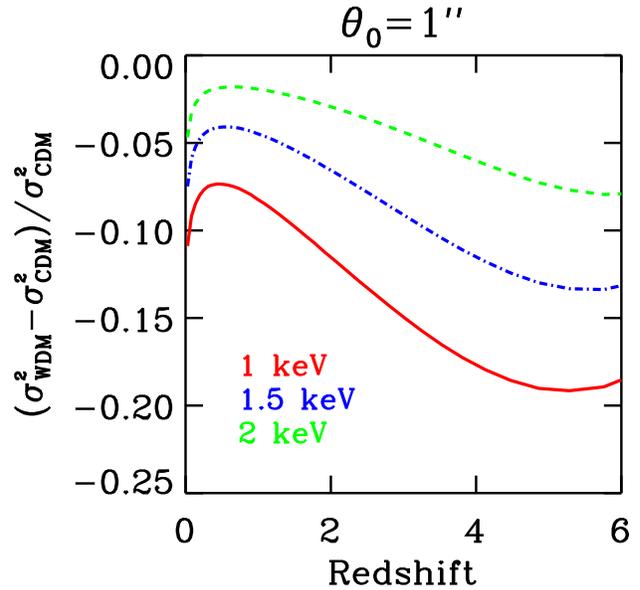}
\caption{Relative difference of the two-dimensional variance of the WDM projected (two-dimensional) matter overdensity distribution with respect to the CDM case as a function of redshift for three choices of the WDM particle mass, $m_\chi=1, 1.5, 2.0$ keV (red, blue, green, respectively).}
\label{fig:sigma}
\end{figure}
 
In this work, we propose a new method to put constraints on WDM particle mass based on strong lensing of high-$z$ SNe due to cosmological matter density fluctuations. As we will see, this probe is very sensitive to the power spectrum of matter along the line of sight, encoding information on the putative WDM particle mass. Moreover, gravitational lensing methods do not rely on assumptions about the coupling between dark and luminous objects~\citep{Markovic:2013}, 
as they directly probe the total gravitational potential. We will compute the strong lensing probability for a distant SN (that is, the probability for a distant SN to undergo a strong lensing event), specifically predicting double or multiple images. This quantity will be then used to discriminate between CDM and WDM cosmologies. The key idea is to isolate the expected differences in the abundance of lenses predicted by these two models. As SNe are very bright sources, they can be detected at medium-high redshifts. It is then possible to investigate a large part of the history of the Universe, up to redshifts at which  small-scale structure is exponentially suppressed by the presence of WDM. 

Strong lensing has been already used in the past to constrain the WDM particle mass. In fact, the intergalactic haloes in the mass range $10^6-10^8$~M$_\odot$ can produce multiple images of distant quasars, as investigated by~\citet{Xue:2001}, but also can induce modifications in the fluxes of the QSO multiple images, as proposed by~\citet{Miranda:2007}. The number of the objects in this mass range is sensitive to differences in the initial power-spectrum at scales useful to discriminate between different warm candidates. 
The main difference between their approach and the present one is that we are interested on scales much larger than the typical halo ones. For this reason. a detailed modelling of the subhalo structure is not necessary.

The paper is organized as follows: in Section~\ref{method}, we explain how we compute the strong lensing probability for a distant source in a given W/CDM cosmology; in Section~\ref{results}, we convolve this result with the cosmic SN rate to obtain the frequency of multiply imaged SNe. The minimum observing time for an SN experiment with a given field of view (FOV) required to disentangle CDM from the lightest WDM particle is derived and discussed. Finally, conclusions are given in Section~\ref{sec:conc}.

\begin{figure}
\centering
\includegraphics[width=8.5cm]{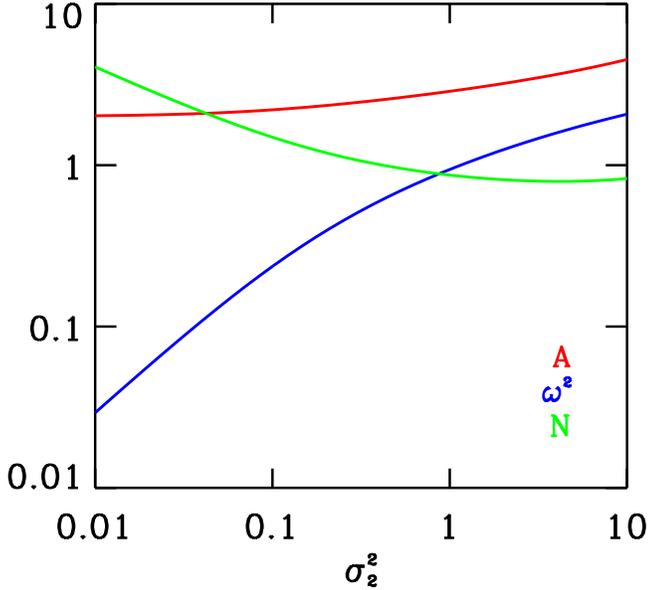}
\caption{Values of parameters $A$, $\omega^2$ and $N$ as a function of $\sigma^2_2(x)$ as indicated. The values of the fitting function parameters are function only of the absolute value of $\sigma_2^2$, and therefore the dependence on the cosmological model (WDM versus CDM) is implicitly carried by the dependence on the $\sigma_2^2$ value.}
\label{fig:params}
\end{figure}

\section{Method}\label{method}

We compute the probability of multiple imaging of distant SNe following the method of \cite{Das:2005}. 
\begin{figure*}
\centering
\includegraphics[width=8.7cm]{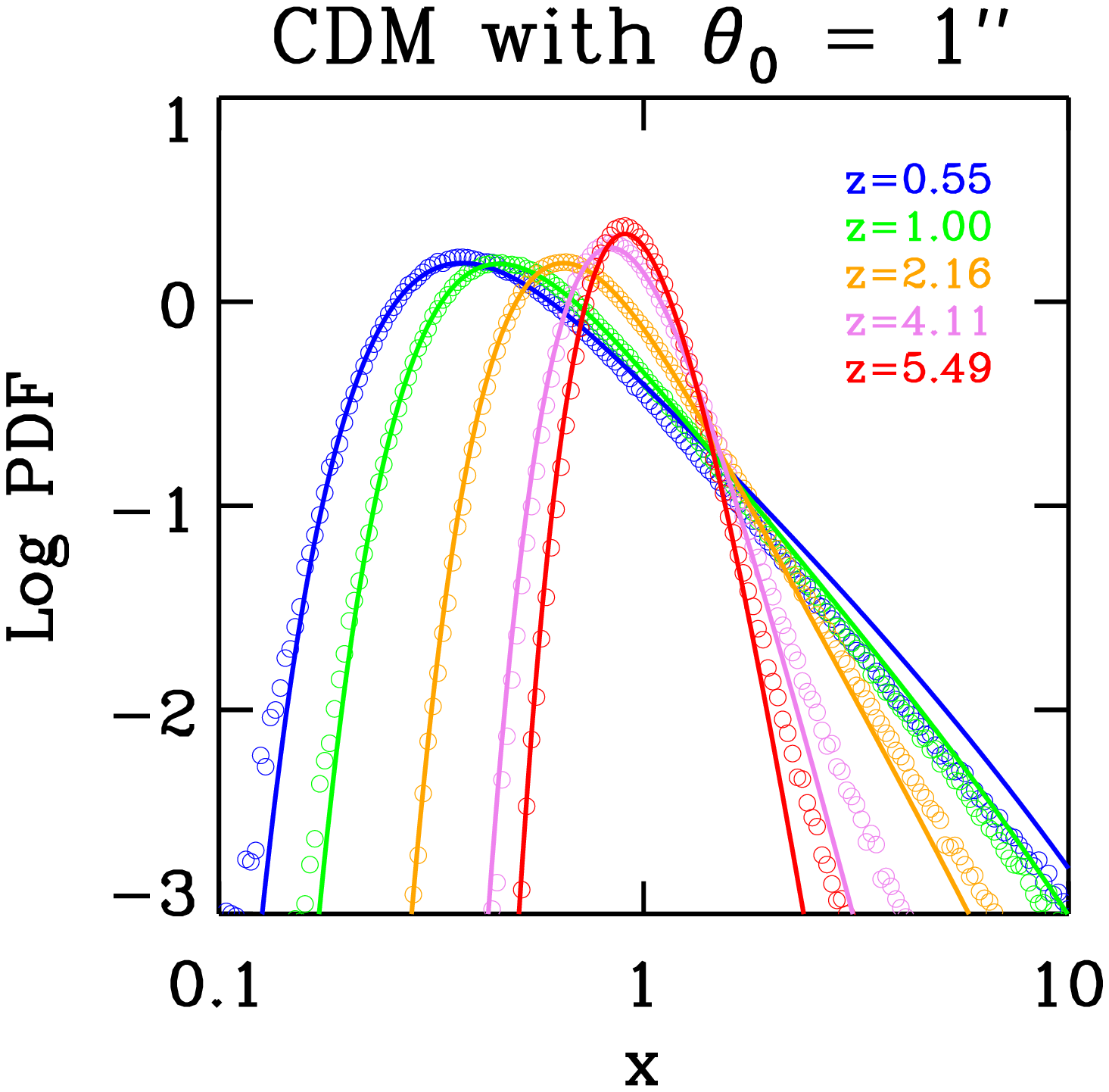}
\includegraphics[width=8.7cm]{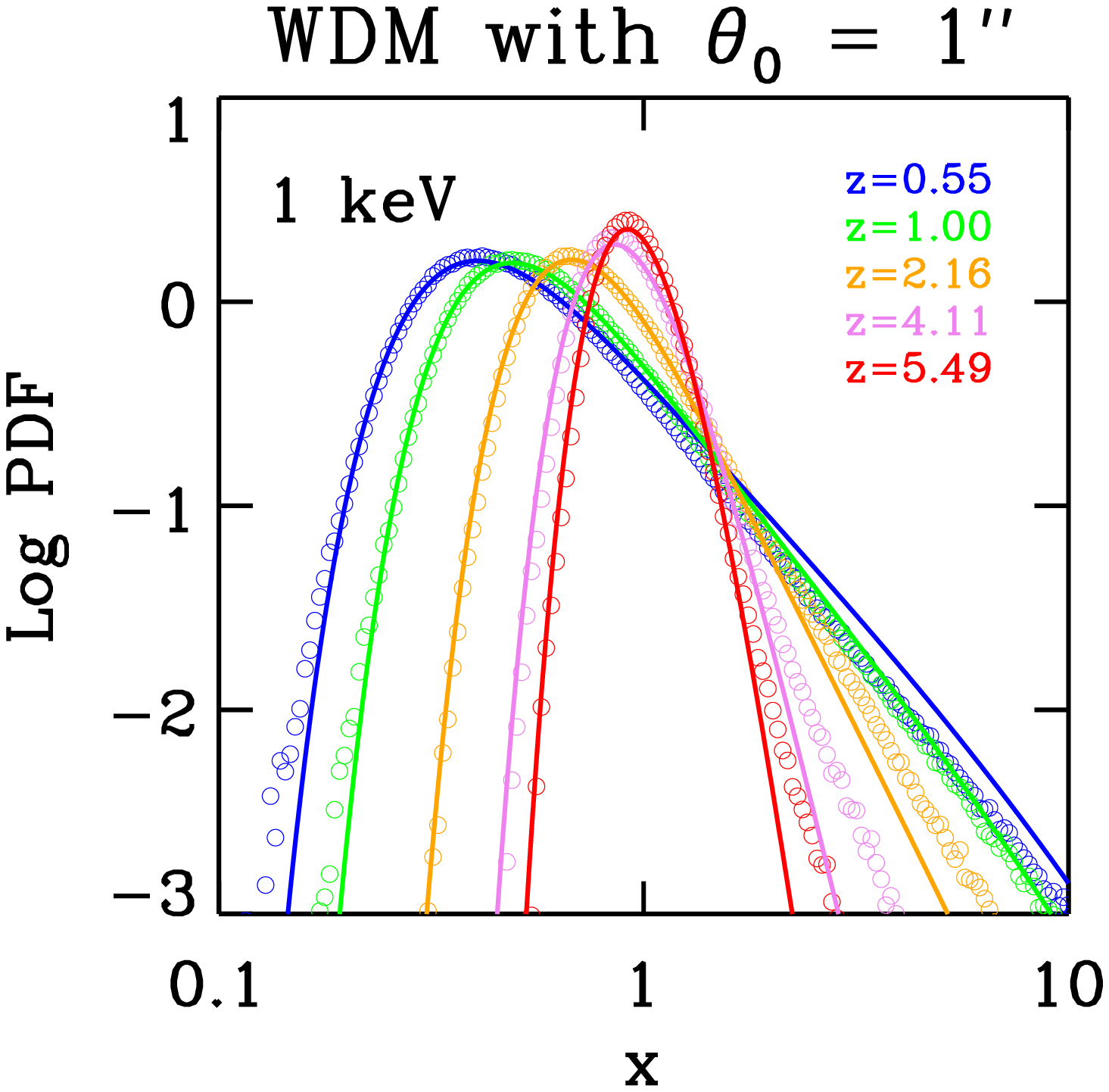}
\caption{Comparison between mass surface density PDFs obtained for our analytical model (solid lines) and those obtained from numerical simulations (circles) described in the text for different redshifts.}
\label{fig:fittingfunction}
\end{figure*}
To this aim, we consider the redshift range $0<z<6$, and divide it into 38 slices $\chi_i, i=0,...,37$, each 160 $h^{-1}$Mpc thick. 
Following \citet{Das:2005}, we assume null covariance between contiguous slices, as the thickness of each slice is chosen to be larger than the correlation length. 

We compute the 3D non-linear matter power spectrum in each slice by using the HALOFIT routine of~\cite{Smith:2003};
 such spectrum is then projected along the line of sight from us to the source on to the central plane of each slice, by using the Limber approximation~\citep{Kaiser:1998}.
 We then extract the variance of the projected (two-dimensional) matter overdensity distribution on each slice $\sigma_2^2(\theta_0,z_i)$, by assuming a Gaussian window of angular radius $\theta_0$:
\beq
\label{eq:sigma}
\sigma_2^2(\theta_0,z_i)=\frac{\pi}{(\chi_i-\chi_{i-1})^2}   \int^{\infty}_0 d l\: {\rm e}^{- {l^2 \theta_0^2}}  \int_{ l/\chi_i}^{l/\chi_{i-1}}\frac{{\rm d} k}{k^3}  \dnl^2,
\eq
where $\dnl^2=(2\pi^2)^{-1}\pnl(k) k^3 $ is the dimensionless power spectrum as a function of scale $k$ and redshift $z$ computed using HALOFIT, $l=k/\chi$ and  ${\rm e}^{- {l^2 \theta_0^2}}$ is the Fourier transform of the Gaussian smoothing window of width $\theta_0$.  The matter power spectrum is computed at the redshift corresponding to the centre of each slice. 
The $\Lambda$CDM cosmological model assumed is given by the marginalized value of the cosmological parameter of WMAP9~\citep{Hinshaw:2013}: 
$\Omega_{\rm m}=0.279$, $\Omega_\Lambda=0.721$, $\Omega_{\rm b}=0.0463$, $h=0.70$, $n_s=0.972$ $\sigma_8=0.821$. We make use of the most updated cosmological parameters values, but we checked that small variations around this choice have a negligible effect on our final results.

In order to see how a WDM scenario would change the matter power spectrum and, therefore, the probability that a distant source undergoes a strong lensing event, we repeat the same procedure described above, for a cosmology with WDM, instead of CDM.

As for the WDM case, we follow the work of \cite{Viel:2012} 
in which the authors run cosmological $N$-body simulations for several WDM cases and eventually provide a fitting formula for the non-linear transfer function ($T_{nl}$):
\begin{eqnarray} \label{eq_fitting}
&&T^2_{\rm nl}(k)\equiv P_{\rm WDM}(k)/P_{\rm \Lambda CDM}(k)=(1+(\alpha\,k)^{\nu l})^{-s/\nu} \nonumber , \\
&& \alpha(m_{\rm WDM},z)=0.0476\ \left(\frac{1 \rm{keV}}{m_{\rm WDM}}\right)^{1.85}\,\left(\frac{1+z}{2}\right)^{1.3} \ ,
\end{eqnarray}
with $\nu=3$, $l=0.6$ and $s=0.4$; $\alpha$ defines the cutoff scale beyond which the power spectrum is exponentially suppressed by free streaming of WDM particles. 
The $N$-body simulations of~\cite{Viel:2012} 
assume WDM particles to be thermal relic fermions; the initial conditions of these simulations are set to reflect the small-scale suppression of the power spectra due to the WDM particle as described in~\cite{Viel:2005}. 
The above fitting formula allows us to derive the non-linear matter power spectrum of $\Lambda$WDM case once the analogous $\Lambda$CDM one is known (from, e.g., HALOFIT).

Fig.~\ref{fig:sigma} shows the difference in the variance of the two-dimensional matter power spectrum $\sigma_2^2(z)$ between WDM and CDM as a function of redshift. This is computed according to equation~(\ref{eq:sigma}) for a smoothing angle $\theta_0=1$ arcsec and three different choices of the WDM particle mass,  $m_\chi=1, 1.5, 2.0$ keV.  We choose the value $\theta_0=1$ arcsec for two main reasons: (i) it corresponds to the smallest scale at which the analytical probability distribution function (PDF) model of~\cite{Das:2005} 
is a good approximation of numerical simulations on non-linear scales; (ii) given the condition (i), $\theta_0=1$ arcsec is the value maximizing our signal, i.e.~is the scale at which we expect the largest differences (in terms of $\sigma_2^2(z)$) between the CDM and WDM. Indeed, for larger, linear, angular scales the relative difference between CDM and WDM tends to vanish. As expected, smaller WDM particle masses result in a lower variance of the two-dimensional matter power spectrum.

To compute the PDF of strong lensing events we again follow the work of~\cite{Das:2005}. 
They propose a phenomenological analytical formula, inspired by the log-normal distribution  for the surface mass density, in terms of the variable $x=\Sigma/\langle \Sigma \rangle$, i.e. the ratio between the projected surface mass density $\Sigma$, smoothed with a Gaussian window of angular radius $\theta_0$,  and the background matter density $\langle \Sigma \rangle$:
\begin{equation}\label{mln}
f(x)=\frac{N}{x} \exp \left[ -\frac{ (\ln x+\omega^2/2)^2(1+ A/x) }{ 2 \omega^2 } \right] \, .
\end{equation}

The values of the three parameters, $A$, $\omega^{2}$ and $N$ are fixed by the following three constraints:
\beqn
\label{constr1}
 \int_{0}^{\infty} f(x) dx& = &1,\\
\label{constr2}
 \int_{0}^{\infty} x f(x) dx& =& 1, \\
\label{constr3}
\int_{0}^{\infty} (x-\langle x\rangle)^2 f dx&=&\sigma_2^2. 
\eeqn
Equation~(\ref{constr1}) represents the normalization of the distribution; equation~(\ref{constr2}) accounts for the fact that, by definition, $\langle x \rangle=1$ and equation~(\ref{constr3}) is the variance constraint that comes from the assumed cosmological model.
The system is solved numerically via a searching algorithm. Once the values of $A$, $\omega^{2}$ and $N$ are obtained, the PDF is completely determined by the variance of the projected surface mass overdensity on each slice, $\sigma^2_2(\theta, z)$. We plot in Fig. \ref{fig:params} the three parameters as a function of $\sigma^2_2$;  note that the chosen cosmology enters here only through the value of $\sigma^2_2(\theta, z)$. This implicitly assumes that the functional shape of the PDF (equation~\ref{mln}) is essentially the same for WDM and CDM provided that the two models are normalized to the same \sdz value. This is a critical issue and we have paid special attention to clarify it, as explained next. 

To this aim, we make use of $N$-body simulations for three different cosmological models: a CDM model, and two WDM models, with particle mass of 1 and 2 keV. In all the three cosmological models, the initial conditions have been generated at $z=99$ by displacing the positions of the particles, that were set into a regular cubic grid, by using the Zel'dovich approximation. The size of the periodic simulation box is 50 $h^{-1}$Mpc, and the number of CDM/WDM particles is equal to $512^3$. The Plummer equivalent gravitational softening is set to 2.5 $h^{-1}$kpc. The cosmological parameters of the three different cosmologies are as follows: $\Omega_{\rm m}=0.2711$, $\Omega_\Lambda=0.7289$, $\Omega_{\rm b}=0.0463$, $h=0.703$, $n_s=0.964$ $\sigma_8=0.809$\footnote{These values are different from the ones we use along the rest of the paper. In fact, purely for comparison sake,  we recompute the $\sigma^2_2$ using the same cosmological parameters as the ones used for the $N$-body simulations.}.
The $N$-body simulations have been run with the GADGET-3 code, which is an improved version of the code Gadget-2 ~\citep{Springel:2005}. The initial matter power spectrum and the transfer functions have been computed by using CAMB~\citep{Lewis:2002}. 
For the WDM models we have used the transfer functions presented in \cite{Bode:2000}. 
We have assumed that the WDM consists in particles whose momentum distribution follows the Fermi-Dirac distribution.

We proceed in the following way, complying with~\cite{Das:2005}: for a given $N$-body snapshot, we project all the particle positions into the $XY$ plane. We then compute the values of surface density, $\Sigma(\vec{r})$, into a grid with $1024^2$ points using the Cloud-in-Cell interpolation procedure. We use the fast Fourier transform (FFT) to calculate the values of the surface density in Fourier space, $\Sigma_k(\vec{k})$. The smoothing of the surface density field is performed by multiplying $\Sigma_k(\vec{k})$ by $\exp(-(kR)^2/2)$, where $k=|\vec{k}|$ and $R$ is the smoothing radius. The relationship between the $R$ and $\theta_0$ is given by $R(z)=D_c(z)\theta_0$, being $D_c(z)$ the comoving distance to redshift $z$. Finally, we FFT back the convolved $\Sigma_k(\vec{k})$ field to obtain the smoothed surface mass density field in real space: $\Sigma_\theta(\vec{r})$. We compute the values $\sigma_2^2(\theta_0,z)$ by 
\begin{equation}
\sigma_2^2(\theta_0,z)=\left\langle \left(\frac{\Sigma_\theta(\vec{r},z)-\langle\Sigma_\theta(\vec{r},z)\rangle}{\langle\Sigma_\theta(\vec{r},z)\rangle} \right)^2 \right\rangle
\end{equation}
In Fig. \ref{fig:fittingfunction} we show with open circles the values of the $\Sigma_\theta(\vec{r},z)$ PDF computed from the $N$-body simulations together with the function $f(x)$ whose parameters have been computed by requiring that it fulfils equations~(5)-(7).
As seen from Fig.~\ref{fig:fittingfunction}, the analytical fitting function for the PDF of the two-dimensional smoothed mass overdensity is in good agreement with both CDM and WDM simulations. Therefore, our analytic method to model lensing can accurately compete  with expensive ray shooting simulations for either CDM or WDM cosmologies. 

From the surface mass density PDF, we can get the PDF for the convergence, $\kappa$, that is the PDF of the surface mass density scaled by the critical (surface) mass density. We first define the lensing surface mass density: 
\beq
\label{sigmalens}
\Sigma_l\equiv\Sigma-\langle \Sigma\rangle.
\eq
The condition for a source to have multiple images, in a single plane case, is that its ray bundles encounter a region whose $\Sigma_l > \Scr$, where the critical surface mass density at redshift $z$ is given by
\beq
\Scr(z)=\frac{c^2}{4 \pi G} \frac{D_s}{D_{ls} D_l}, 
\eq
where $D_s$, $D_l$ and $D_{ls}$ are, respectively, the angular diameter distances from the observer to the source, from the observer to the lens, and from the lens to the source. We can finally switch to convergence 
\beq
\label{kappa}
 \kappa=\frac{\Sigma_l}{\Scr}=\frac{\langle \Sigma \rangle}{\Scr} (x-1).
\eq
The convergence PDF is simply related to the PDF of the surface mass density by the Jacobian transformation 
\beq 
g(\kappa)=\alpha f(\alpha \kappa+1)
\eq
with $\alpha=\Scr/\langle \Sigma \rangle$.

\begin{figure}
\centering
\includegraphics[width=8.5cm]{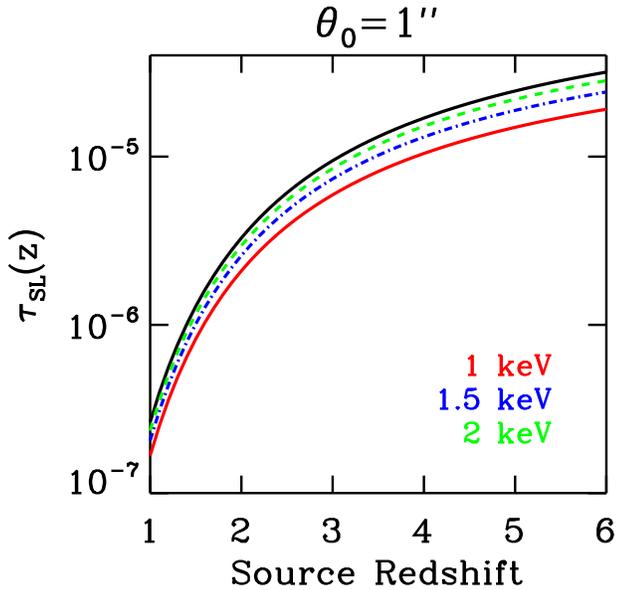}
\caption{Probability of strong lensing for a source at a given redshift, for the different cosmological models. The (black) continuous line at the top is the CDM cosmology case; cosmologies with a different WDM particle mass  $m_\chi=1, 1.5, 2.0$ keV are shown by red (solid), blue (dot-dashed), and green (dashed) lines, respectively.}
\label{fig:prob}
\end{figure}

We randomly draw values $\kappa_i$ from the convergence PDF for every plane in front of the source. The multiple imaging condition, extended to the case of multiple plane case, due to strong lensing events is $\sum \kappa_i>1$~\citep{Das:2005}. 
For this reason, to calculate the probability for a source at redshift $z$ of undergoing strong lensing event (also called the strong lensing optical depth of $\tau_{SL}(z)$), we evaluate how many times the total convergence is $>1$ over 1000 realizations.
Even if the above condition $\sum \kappa_i>1$ might in principle be given by more than one plane (i.e. when the largest value of the convergence is not already higher than one ($\kappa_{max}>1$), but the condition  $\sum \kappa_i>1$ is satisfied by two or more planes), we checked that the probability of a strong lensing event produced by two or more planes is $\simlt 10\%$. In other words, this means that in the majority of the cases the strong lensing event is produced by a single plane (see also discussion in section~4 of~\citealt{Das:2005}).

We plot the strong lensing optical depth in Fig.~\ref{fig:prob} for three WDM particle masses as a function of redshift. The difference with the CDM case is of the order of 30\% for the 1 keV particle at redshift $z=5$.
The differences decrease at lower redshift and for higher WDM particle masses. Therefore, at redshift $z=4$, in order to see at least one strong lensing event it is necessary to observe $\approx 10^5$ SNe if the underlying cosmological model has a WDM particle mass of 1 keV; while instead for the CDM cosmology, to have a strong lensing event we would need to observe a 40\% less SNe, as the probability of strong lensing is higher.

\section{Analysis and results}\label{results}

Given the probability of strong lensing for distant sources in different cosmological scenarios, we can convolve it with the rate of SNe explosions as a function of redshift, in order to obtain the main result of our study: the rate of multiply imaged SNe. 
As already mentioned, this quantity is sensitive to the underlying cosmology and it could be used to put constraints on the mass of the WDM particles.

\begin{figure}
\centering
\includegraphics[width=8.5cm]{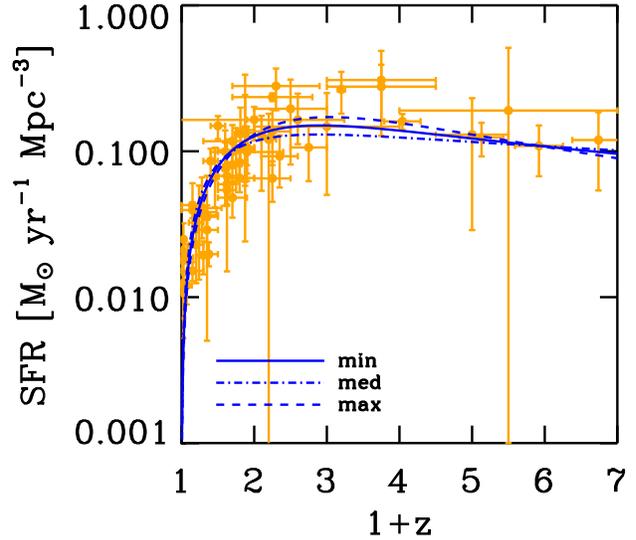}
\caption{The SFR from the observations collected in~\citet{Hopkins:2006} fitted with our function with the parameters in Table \ref{tbl:fit} for the best-fit (solid line) and 3$\sigma$c.l. region around the best-fit.}
\label{fig:SFR}
\end{figure}

The rate of the lensed SNe up to redshift $z$ is
\begin{equation}
\dot{N}_{\rm SN}(z) = \int_{0}^{z} dV_{c}(z') \tau_{SL}(z) R_{\rm SN}(z')
\label{SNrate}
\end{equation}
where $V_{c}(z)$ is the comoving volume explored by the survey, and $\tau_{SL}(z)$ is the probability of an SN  to undergo a strong lensing event. The SN rate, $R_{\rm SN}$, depend on cosmic star formation history (SFH), $\psi(z)$, and stellar initial mass function (IMF) $\phi(m)$. We assume a Salpeter IMF function with $x=-1.35$ and $m_{\rm cut} = 0.35$~M$_\odot$ in the range $[0.1,100]$~M$_\odot$ as derived in~\citet{Larson:1998}. The frequency of core-collapse SNe, SNe II and possibly SNe Ib/c, which have short-lived progenitors is essentially proportional to the instantaneous stellar birthrate of stars with mass $>8~M_\odot$:

\begin{equation}
R_{\rm SN\,II}(z) = \psi(z) \frac{\int_{8}^{50} dm \, \phi(m)}{\int_{0.1}^{100} dm \, m \, \phi(m)}.
\end{equation}
For Type Ia SN (SN Ia) we account for the delay time $t_{\rm Ia}$ between progenitor formation and the SN event, further assuming a 
fraction $\eta$ stars in the mass range $3~M_\odot <M< 8~M_\odot$ to end their lives as SN Ia. We then write the rate as
\begin{equation}
R_{\rm SN\,Ia}(z) = \eta \psi(t-t_{\rm Ia}) \frac{\int_{3}^{8} dm \, \phi(m)}{\int_{0.1}^{100} dm \, m \, \phi(m)} \, ,
\end{equation}
with $\eta = 0.05$ and $t_{\rm Ia} = 1$~Gyr following \citet{Hopkins:2006}. 
The SFH is taken from the collection of observation reported in \citet{Hopkins:2006}; we fit the data with a function similar to the one proposed by \cite{Cole:2001}
 with three free parameters: $f(z) = bz/[1+(z/c)^d]$. The fitting is obtained by a $\chi^2$ minimization to $50$ selected measurements of the SFH spanning from $0\lsim z \lsim 6$. We obtain a best fit of $\chi^2=46.7$, without assuming any covariance among the data. In order to estimate the uncertainty, we compute the best-fitting SFH from \citet{Hopkins:2006}
 data and then we determine all the SFHs within 3$\sigma$ c.l. around the best fit of the SFH at $z=3$, i.e. the SFH peak redshift. The value corresponding to $\pm 3\sigma$ are considered as the minimum and maximum SFH allowed at that redshift. The best-fitting values for $b, c, d$, along with one, two and three $\sigma$ c.l. are given in Table~\ref{tbl:fit} for $z=3$ (see Fig. \ref{fig:SFR}).

\begin{figure}
\centering
\includegraphics[width=8.5cm]{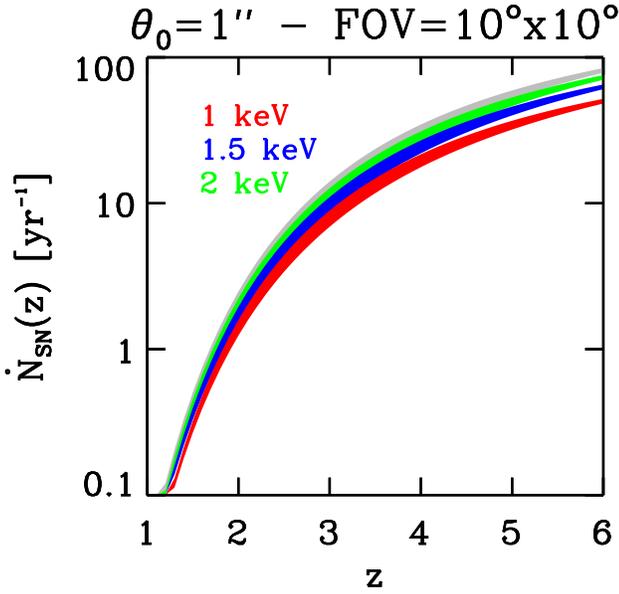}
\caption{Rate (in yr$^{-1}$) of lensed SNe up to redshift $z$ for an FOV of $100$~square degrees. The shaded region accounts for the SFH uncertainties. The grey area corresponds to the CMD case.}
\label{fig:NSN}
\end{figure}

\begin{figure}
\centering
\includegraphics[width=8.5cm]{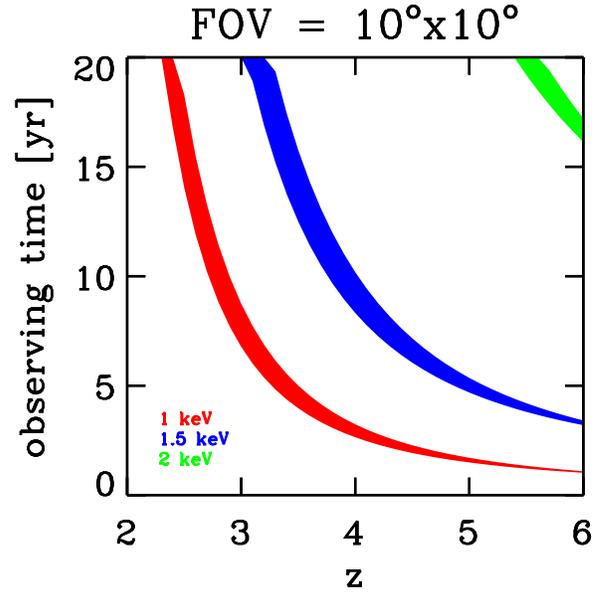}
\caption{Minimum required observing time for a 2$\sigma$ c.l. discrimination between a CDM model and a WDM scenario, as a function of redshift and WDM mass for a future survey with an FOV of $100$~square degrees.}
\label{fig:time}
\end{figure}

We plot in Fig.~\ref{fig:NSN} the expected rate of strongly lensed SNe (i.e. multiply imaged SNe) as function of the maximum redshift $z$ reached by a future survey with an FOV of $100$~square degrees. We have taken into account the uncertainty in the determination of the SFH, as explained above, and used it to assign an error on the number of lensed SN,

As the relative difference in the number of lensed SNe among different dark matter cases is relatively small, it is important to derive the minimum observing time required to disentangle a given WDM scenario from the standard CDM one at a certain c.l. 
We proceed as follows. Given the rate $\dot{N}_{\rm SN, i}(z)$ of lensed SNe, up to redshift $z$, for an assumed cosmological model $i={\rm CDM, WDM}$, we can associate a Poissonian error at a given c.l: after $t_{obs}$ years we expect to detect, say at 2$\sigma$ c.l., a maximum (minimum) number $N_{\rm SN, CDM} + 2 \sqrt{N_{\rm SN, CDM}}$ ($N_{\rm SN, WDM} - 2 \sqrt{N_{\rm SN, WDM}}$) of lensed SNe,  where $N_{\rm SN, i} (z)= \dot{N}_{\rm SN, i}(z)t_{obs}$, and $\dot{N}_{\rm SN, i}(z)$ is given by equation~(\ref{SNrate}).

We want to distinguish the CDM from the WDM cosmology only by counting the number of strong lensed SNe, and comparing it with the number expected from the corresponding theoretical model. Therefore, the minimum observing time required to disentangle at 2$\sigma$ a CDM cosmology from a 1 keV WDM particle cosmology, is given by the condition $N_{\rm CDM} - 2 \sqrt{N_{\rm CDM}} > N_{\rm WDM} +  2 \sqrt{N_{\rm WDM}}$. This condition is shown in Fig. \ref{fig:time}. From there we conclude that in 2 yr, a survey with an FOV of 100 square degrees sensitive to SNe up to $z=5$, will be able to discriminate between a WDM model with $m_\chi=1$ keV and a CDM model.

Upcoming (SN) surveys like Euclid~\citep{Amendola:2013}, and the extremely large telescopes (ELT) (such as the European-ELT (E-ELT)
\footnote{http://www.eso.org/sci/facilities/eelt/}) 
are going to provide a large gain both in terms of number of SN observed and survey depth \citep{Hook:2012}.
Euclid's wide field survey will cover 15000 square degrees, with limiting magnitude AB = 24.5 in a single band (R+I+Z); in addition, a deep field of $\approx$ 40 square degrees will go about 2 mag deeper; its expected lifetime is of 6 yr. According to simulations, it will observe 5000 SNe Ia up to redshift $z=1.5$~\citep{Hook:2012}. 
The E-ELT is a 39m diameter optical-IR telescope, which will see its first light in 2022. According to the E-ELT Science Case \citep{Hook:2005}, 
the expected number of SNe in an FOV of 2 arcmin squared will be 4-7 SNe of Types Ia, Ib/c and II per field yr$^-1$ up to $z\sim 5-8$, by taking four exposures at time intervals of 3 months of 50 fields in the J, H, and K bands of 1 h each. Moreover, there is planned a photometric follow-up of about 350 SNe up to $z\sim 10$. Therefore, in about 4 months of total time of observations, E-ELT will be able to study 400 SNe up to redshift $z\sim$ 10~\citep[see][and reference therein]{Hook:2005}. 

While discriminating CDM from WDM appears very challenging given the current situation, some of the planned future experiments, such as the E-ELT, maybe crafted to consider the needed experimental specifications, particularly in terms of survey sensitivity.

\begin{table}
\begin{center}
\label{tbl:fit}
\begin{tabular}{l c c c c c c c}
\hline
\hline
Parameter & best-fit & \multicolumn{2}{c}{$1\sigma$} & \multicolumn{2}{c}{$2\sigma$} & \multicolumn{2}{c}{$3\sigma$} \\   
          &          & min & max & min & max & min & max \\   
\hline
b\dotfill & 0.25 & 0.27 & 0.23 & 0.28 & 0.22 & 0.30 & 0.21 \\
c\dotfill & 1.73 & 1.45 & 1.97 & 1.35 & 2.11 & 1.17 & 2.26 \\
d\dotfill & 1.83 & 1.66 & 2.01 & 1.58 & 2.13 & 1.49 & 2.26 \\
\hline 
\hline 
\end{tabular}
\caption{Best-fitting values for the SFH fitting function parameters $b, c, d$, along with their 1$\sigma$, 2$\sigma$ and 3$\sigma$~c.l.~for $z=3$.}
\end{center}
\end{table}

\section{Conclusions}
\label{sec:conc}
We have proposed a new method to constrain WDM candidate particle mass, based on the counts of multiply imaged, distant SNe produced by their strong lensing by cosmological matter fluctuations. 
We have extended the \cite{Das:2005} analytical model for the probability density function of matter fluctuations for the WDM case, and tested it against $N$-body simulations for WDM cosmologies. We compute the expected number of strongly lensed  high-$z$ SNe for the standard $\Lambda$CDM model and three variants of the WDM model, differing for the particle mass ($m_\chi=1, 1.5, 2$ keV). At redshift $z=4$, to see one strong lensing event requires the observation of $\approx 10^5$ SNe (WDM, particle mass of 1 keV)  or $\approx 6 \times 10^4$ SNe (CDM). A minimum observing time of 2 yr (5 yr) is required for a future 100 square degrees survey reaching $z \gtrsim 4$ ($z \gtrsim 3$) are needed to disentangle at 2$\sigma$ a WDM ($m_\chi=1$ keV) model from the standard CDM scenario.

This method is not affected by any astrophysical uncertainty and, in principle, it does not need a dedicated strategy survey, as it can come as a byproduct of a future SN surveys. Moreover, as lensing directly probes the total gravitational potential of the intervening matter, the present method is free from complex baryonic physics effect that usually affect other types of experiments, such as the Lyman-$\alpha$ forest~\citep{Viel:2008} or the reionization data~\citep{Schultz:2014}.

Future surveys (Euclid and E-ELT) specifically designed to observe SNe at redshifts higher than $z \gtrsim 3$ may put alternative and robust constraints on the WDM scenario through the proposed experiment. In particular, E-ELT will be able to see a total of 400 SNe of Types Ia, Ib/c and II up to redshift $z\sim$ 10 in about 4 months of observing time~\citep{Hook:2005}.

Beyond the quantitative specific result, an important aspect of this work consist in the extension of the analytical method to compute the (strong) lensing PDF of \citet{Das:2005} 
to the WDM cosmology case. As we already noticed, this analytical method is able to compete with computationally-expensive, time-consuming ray-tracing techniques: we have expanded it to the WDM cosmology case, and we have shown that it is in very good agreement with the corresponding WDM $N$-body simulations. We can therefore use this theoretical machinery to further investigate the lensing properties of a WDM Universe. 

\section*{Acknowledgements}
We would like to thank Matteo Viel, Alberto Vallinotto and Sherry Suyu for useful discussions. 
The Dark Cosmology Centre is funded by the Danish National Research Foundation. CE acknowledges support from the ''Helmholtz Alliance for Astroparticle Physics HAP'' funded by the Initiative and Networking Fund of the Helmholtz Association. FVN is supported by the ERC Starting Grant ''cosmoIGM''.

\bibliographystyle{mn2e}
\bibliography{WDM}
\end{document}